\documentclass[acmtog]{acmart} 

\acmJournal{TOG}

\usepackage{booktabs} 
\setcitestyle{square}
\usepackage{ifthen}
\usepackage{enumitem}
\usepackage{algorithm}
\usepackage{algorithmicx}
\usepackage[noend]{algpseudocode}
\usepackage{syntax}
\usepackage{amsfonts}
\usepackage{listings}
\usepackage{fancyvrb}
\usepackage{wrapfig}
\usepackage{graphicx}
\usepackage{subcaption}
\usepackage{xspace}
\usepackage{comment}
\usepackage{hyperref}
\usepackage{multirow}
\usepackage{xfrac}

\usepackage{units}

\setlist{leftmargin=*} 

\algdef{SE}[DOWHILE]{Do}{doWhile}{\algorithmicdo}[1]{\algorithmicwhile\ #1}%

\usepackage{algorithm}  
\usepackage{algpseudocode}  
\usepackage{amsmath}  

\usepackage{colortbl}
\makeatletter
\newcommand{\oset}[3][0ex]{%
  \mathrel{\mathop{#3}\limits^{
    \vbox to#1{\kern-2\ex@
    \hbox{$\scriptstyle#2$}\vss}}}}
\makeatother

\definecolor{mygreen}{rgb}{0,0.6,0}                                             
\definecolor{mygray}{rgb}{0.5,0.5,0.5}                                          
\definecolor{codebg}{rgb}{1, 1, 0.9}  
\usepackage[normalem]{ulem}

\newcommand{\inchsign}{^{\prime\prime}}  
\renewcommand{\grammarlabel}[2]{#1\hfill#2}

\begin{document}
\title{Co-Optimization of Design and Fabrication Plans for Carpentry: Supplemental Material}

\author{Haisen Zhao}
\email{haisen@cs.washington.edu}
\affiliation{%
  \institution{University of Washington and Shandong University}
}

\author{Max Willsey}
\email{mwillsey@cs.washington.edu}
\affiliation{%
  \institution{University of Washington}
}

\author{Amy Zhu}
\email{amyzhu@cs.washington.edu}
\affiliation{%
  \institution{University of Washington}
}

\author{Chandrakana Nandi}
\email{cnandi@cs.washington.edu}
\affiliation{%
  \institution{University of Washington}
}

\author{Zachary Tatlock}
\email{ztatlock@cs.washington.edu}
\affiliation{%
  \institution{University of Washington}
}

\author{Justin Solomon}
\email{jsolomon@mit.edu}
\affiliation{%
  \institution{Massachusetts Institute of Technology}
}

\author{Adriana Schulz}
\email{adriana@cs.washington.edu}
\affiliation{%
  \institution{University of Washington}
}

%
%

\newcommand{\hsyntax}[1]{\ensuremath{\mathit{#1}}}
\renewcommand{\grammarlabel}[2]{#1\hfill#2}

\newcommand{\adriana}[1]{{\bfseries \scriptsize \textcolor[rgb]{0.00,0.70,0.00}{AS: #1}}}
\newcommand{\zach}[1]{{\bfseries \scriptsize \color{red} ZT: #1}}
\newcommand{\justin}[1]{{\bfseries \scriptsize \color{blue} JS: #1}}
\newcommand{\amy}[1]{{\bfseries \scriptsize \textcolor[rgb]{0.80,0.00,0.80} {AZ: #1}}}
\newcommand{\haisen}[1]{{\bfseries \scriptsize \color{red} HZ: #1}}
\newcommand{\chandra}[1]{{\bfseries \scriptsize \color{purple} CN: #1}}
\renewcommand{\max}[1]{{\bfseries \scriptsize \color{orange} MW: #1}}

\newcommand{\change}[1]{{\color{black}#1}}

\newcommand{\Egraph}{{E-graph}\xspace}
\newcommand{\Egraphs}{{E-graphs}\xspace}
\newcommand{\egraph}{{e-graph}\xspace}
\newcommand{\egraphs}{{e-graphs}\xspace}
\newcommand{\eclass}{{e-class}\xspace}
\newcommand{\eclasses}{{e-classes}\xspace}
\newcommand{\Eclasses}{{E-classes}\xspace}
\newcommand{\enode}{{e-node}\xspace}
\newcommand{\enodes}{{e-nodes}\xspace}
\newcommand{\Enodes}{{E-nodes}\xspace}
\newcommand{\llhelm}{{LL-HELM}\xspace}
\newcommand{\hlhelm}{{HL-HELM}\xspace}
\newcommand{\fabplan}{{fabrication plan}\xspace}
\newcommand{\fabplans}{{fabrication plans}\xspace}
\newcommand{\tool}{{Carpentry Compiler 2.0}\xspace}
\newcommand{\submodule}{{e-class}\xspace}
\newcommand{\submodules}{{e-classes}\xspace}
\newcommand{\subprogram}{{sub-program}\xspace}
\newcommand{\subprograms}{{sub-programs}\xspace}
\newcommand{\afabvar}{{fabrication arrangement}\xspace}
\newcommand{\afabvars}{{fabrication arrangements}\xspace}
\newcommand{\cutEplan}{{cutting order related plan}\xspace}
\newcommand{\cutEplans}{{cutting order related plans}\xspace}
\newcommand{\term}{\ensuremath{\mathcal{T}}}
\newcommand{\bope}{{BOP E-graph}\xspace}
\newcommand{\uscore}{\ensuremath{I_{score}}\xspace}
\newcommand{\escore}{\ensuremath{E_{score}}\xspace}
\newcommand{\sols}{\ensuremath{\mathcal{S}}\xspace}
\newcommand{\nd}{\ensuremath{|\mathcal{D}|}\xspace}

\newcommand{\fabheur}{Heuristic Driven Fabrication Variations\xspace}

\maketitle

\section{Technique details}
This section provides additional details of our ICEE algorithm for extracting a Pareto front where each solution $s$ represents a (design, fabrication) pair.

\subsection{Generating Design Variants}
Design variants can be generated manually by a user or automatically. In our paper, we implement an automatic method, which includes three steps:
(1) detecting all the assembly connectors between two neighboring parts;
(2) enumerating all candidate connector variants for each connect;
(3) generating design variants by selecting different connector variation of each connector, as shown in Figure 4 of the main paper.

\subsection{Fabrication arrangement generation}

Given a specific design variation,
we use a fabrication arrangement generation algorithm
to update the BOP \Egraph such that the \Egraph encodes more \afabvars.
This algorithm uses two heuristic described as Section 4.3.3.

For an input design variation $d_i$ which consists of parts $p_j,0...n$, this algorithm first groups a library of stock lumbers by their dimensions, e.g., lumber $2 \inchsign \times4\inchsign \times 24\inchsign$, lumber $2 \inchsign \times4\inchsign \times 48\inchsign$, and lumber $2 \inchsign \times4\inchsign \times 96\inchsign$ are grouped together. For the parts assigned to this group, this algorithm starts by packing all the parts on the largest stock lumber of the current group. The packing process will be terminated when (1) the current stock lumber is maximally packed or (2) all parts are packed. In the first case, the packing process can continue by switching to another stock lumber (also the largest one) until the second termination condition is reached. Such a layout process is called a full \textit{Traversal} which traverses all of the parts in a specific order.
Prior work~\cite{wu2019carpentry} has used a similar approach but used the number of  \textit{Traversals} as the termination criterion which prevents them to control the amount of \afabvars to generate with this heuristic-driven method.

\subsection{Cost Metrics}

\label{subsec:costs}

This section describes how we compute the three costs:
 material usage ($f_c$),
 cutting precision ($f_p$), and
 fabrication time ($f_t$).
Our formulae are updated versions of the
  ones used by Wu et al.~\shortcite{wu2019carpentry}. Our key improvements are 1) we include a quantitative evaluation method. Each cost metric is associated with a meaningful unit: material usage ($f_c$) is in dollars, cutting precision ($f_p$) is in inches, fabrication time ($f_t$) is in minutes. 2) we evaluate stock load and unload time to make the result fabrication time much more reasonable.


\subsubsection*{Material Cost}
We compute material cost as:
\begin{align*}
    f_c = \sum_{i = 1}^{n} p_i.
\end{align*}
where $p_i$, the price of the $i^{th}$ piece of
 stock is
 estimated based on costs from
 standard US vendors~\cite{mcmaster} as
 shown in \autoref{tab:stockcost}, and $n$
 is the total number of pieces of lumber used.

\begin{table}[ht]
    \small
    \centering
    \begin{tabular}{c|c|c}
    \hline
        Stock &  Dimension & Material Cost (\$)  \\
        \hline
         $2 \inchsign \times 2 \inchsign$ & $24 \inchsign$ & 3.0 \\ 
         $2 \inchsign \times 2 \inchsign$ & $48 \inchsign$ & 5.5 \\ 
         $2 \inchsign \times 2 \inchsign$ & $96 \inchsign$ & 10.0 \\ 
         $2 \inchsign \times4\inchsign$ & $24\inchsign$ & 3.0 \\ 
         $2 \inchsign \times 4 \inchsign$ & $48 \inchsign$ & 5.5 \\ 
         $2 \inchsign \times 4 \inchsign$ & $96 \inchsign$ & 10.0 \\ 
         $4\inchsign \times 4\inchsign$ & $24 \inchsign$ & 7.5 \\ 
         $4\inchsign \times 4\inchsign$ & $48 \inchsign$ & 13.75 \\ 
         $4\inchsign \times 4\inchsign$ & $96 \inchsign$ & 25.0 \\ 
         $2\inchsign \times 8\inchsign$ & $24 \inchsign$ & 7.5 \\ 
         $2\inchsign \times 8\inchsign$ & $48 \inchsign$ & 13.75 \\ 
         $2\inchsign \times 8\inchsign$ & $96 \inchsign$ & 25.0 \\ 
         $1/2 \inchsign$ & $12 \inchsign \times 20 \inchsign$ & 5.5 \\
         $1/2 \inchsign$ & $24\inchsign \times 20\inchsign$ & 10.0 \\
         $1/2 \inchsign$ & $48\inchsign \times 36 \inchsign$ & 30.0 \\
         3/4 $\inchsign$ & $12 \inchsign \times 20 \inchsign$ & 7.0 \\
         3/4 $\inchsign$ & $24 \inchsign \times 20 \inchsign$ & 12.0 \\
         3/4 $\inchsign$ & $48 \inchsign \times 36 \inchsign$ & 32.0 \\
    \end{tabular}
    \caption{\change{Prices of stocks}}
    \label{tab:stockcost}
\end{table}

\begin{table*}
    \small
  \begin{minipage}{0.69\textwidth}
  \centering
    \begin{tabular}{c|c|c|c|c}
        \hline
        \multirow{2}{*}{Tool} & \multicolumn{2}{c|}{Full setup (s)} & \multirow{2}{*}{Partial setup (s)} & \multirow{2}{*}{Op Time}\\
        \cline{2-3}
        & Lumber & Plywood & &\\
        \hline
         Chopsaw  & 60 & 60  & 15  & 1 s \\
         Bandsaw  & 20 & 90  & N/A  & 1 inch/s \\
         Jigsaw   & 30 & 60  & N/A  & 1 inch/s \\
         Tracksaw & 180 & 180 & 75 & 4.5 inch/s \\
         Drill    & 20 & 20 & N/A & 0.1 inch (depth)/s
    \end{tabular}
    
    \caption{\change{Fabrication times for different tools. 
    All times are converted to minutes in our results.}}
    \label{tab:time}
  \end{minipage}
  \begin{minipage}{0.3\textwidth}
  \centering
    \begin{tabular}{c|c}
        \hline
         Tool & Operation Error \\
         \hline
         Chopsaw & $\nicefrac{1}{64}^{th}$ of an inch \\
         Bandsaw & $\nicefrac{1}{16}^{th}$ of an inch \\
         Jigsaw & $\nicefrac{3}{16}^{th}$ of an inch \\
         Tracksaw & $\nicefrac{1}{32}^{nd}$ of an inch \\
         Drill & $\nicefrac{1}{32}^{nd}$ of an inch
    \end{tabular}
    
    \caption{\change{Error per cut for each tool.}}
    \label{tab:prec}
  \end{minipage}
\end{table*}


\begin{table}[ht]
    \small
    \centering
    \begin{tabular}{c|c|c|c|c}
    \hline
        Stock & F-Load (s) & P-Load (s) & F-Unload (s) & P-Unload (s) \\
        \hline
         $2 \inchsign \times 2 \inchsign \times 4 \inchsign$  & 10 & 1 & 5 & 1 \\ 
         $2 \inchsign \times 2 \inchsign \times48 \inchsign$ &20&2&8&2\\ 
         $2 \inchsign \times 2 \inchsign\times96 \inchsign$ & 40 & 3 & 15 & 2 \\ 
         $2 \inchsign \times4\inchsign\times24\inchsign$ & 10 & 1 & 5 & 1 \\ 
         $2 \inchsign \times 4 \inchsign\times48 \inchsign$ & 20 & 2 & 8 & 2 \\ 
         $2 \inchsign \times 4 \inchsign\times96 \inchsign$ & 40 & 4 & 15 & 2 \\ 
         $4\inchsign \times 4\inchsign\times24 \inchsign$ & 15 & 2 & 5 & 1 \\ 
         $4\inchsign \times 4\inchsign\times48 \inchsign$ & 30 & 4 & 10 & 2 \\ 
         $4\inchsign \times 4\inchsign\times96 \inchsign$ & 60 & 6 & 20 & 3 \\ 
         $2\inchsign \times 8\inchsign\times24 \inchsign$ & 15 & 2 & 5 & 1 \\ 
         $2\inchsign \times 8\inchsign\times48 \inchsign$ & 30 & 4 & 10 & 2 \\ 
         $2\inchsign \times 8\inchsign\times96 \inchsign$ & 60 & 6 & 20 & 3 \\ 
         $1/2 \inchsign\times12 \inchsign \times 20 \inchsign$ & 30 & 3 & 10 & 2 \\
         $1/2 \inchsign\times24\inchsign \times 20\inchsign$ & 50 & 5 & 15 & 2 \\
         $1/2 \inchsign\times48\inchsign \times 36 \inchsign$ & 100 & 10 & 20 & 2 \\
         3/4 $\inchsign\times12 \inchsign \times 20 \inchsign$ & 30 & 3 & 10 & 2 \\
         3/4 $\inchsign\times24 \inchsign \times 20 \inchsign$ & 50 & 5 & 15 & 2 \\
         3/4 $\inchsign\times48 \inchsign \times 36 \inchsign$ & 100 & 10 & 20 & 2 \\
    \end{tabular}
    \caption{\change{Loading and unloading time of stocks. All times are converted to minutes in our results.}}
    \label{tab:loadtime}
\end{table}

\subsubsection*{Time}

We asked an expert carpenter to assign a fabrication time
 to each tool (\autoref{tab:time}) and estimate a stock loading and unloading time for each piece of wood stock (\autoref{tab:loadtime}).
They reported the time taken for (1) full setup of the tool,
  (2) partial setup when applicable
  (a partial setup in one where only some parameters are changed),
  (3) full stock load (unload) time
  (setting up a piece of wood stock to the tool workspace)
  (4) partial stock load (unload) time
  (setting up a piece of wood stock to the tool workspace where stocks are stacked)
  and
  (5) performing a single operation (e.g., cut).
Out of all the used tools, tracksaw has the most elaborate setup process
  which is indicated by its long setup times (both full and partial).
Bandsaw and jigsaw are set to different full setup time while cutting lumber and plywood sheet.
Only chopsaw and tracksaw allow partial setups,
  the remaining tools do not.
The operation time for a chopsaw is constant --- all cuts take a second.
For all other tools,
  the operation time is based on the length cut per second.
For example, the operation time for a tracksaw cut of length
$l''$  is $l / 4.5$ seconds.

We refer to a cut as "partial" if it requires only a partial setup.
For example, if the $i^{th}$ cut is a partial cut on a chopsaw,
  the time taken for this cut would be $15s$ for partial setup and
  $1 s$ for the cutting operation leading to a total of $16s$ for the cut to
  be completed.
The fabrication time, $f_t$ is therefore computed as:
\begin{align*}
    f_t = \sum_{i = 1}^{k} (s_i +w_i+ o_i)
\end{align*}
 where $s_i$, $w_i$ and $o_i$ are the setup time, stock load and unload time,
 and operation time for the $i^{th}$ cut respectively, and $k$
 is the total number of cuts.
$s_i$ and $o_i$ are computed based on \autoref{tab:time}. $w_i$ is computed based on \autoref{tab:loadtime}. For a cut with stacking, $w_i$ is measured as the sum of a full load and unload time for the first piece of stock, and multiple partial load and unload times for the following stocks.

\subsubsection*{Precision}
To measure the precision of a fabrication plan,
  we compute (1) measurement error, $\epsilon$, and
  (2) operation error, $p$.
We use $m = 1/16\inchsign$ as the minimum measurement
  that can be made in any of the currently supported tools.
The measurement error for
 length, $m'$ is computed as:
\begin{align*}
 \epsilon = min (m' \% m, m - (m' \% m))
\end{align*}
which, intuitively, is the residual length that cannot be measured
  by our tools.
For operation error,
  we again asked an expert to estimate the error per operation
  for each tool (\autoref{tab:prec}).
The error for the $i^{th}$ cut is therefore
$\epsilon_i + p_i$.
$f_p$ is then computed as:
\begin{align*}
    f_p = \sum_{i = 1}^{k}(\epsilon_{i} + p_i)
\end{align*}
where $k$ is the total number of cuts.
Lower values of $f_p$ indicate higher precision.

\change{
\subsubsection{Mixed-material implementation}
By introducing a new material, we must accommodate the cost of 
using this new material in our metrics. Based on an expert's input, 
we set material price of metal to be 20 times that of wood. All of the same tools can be used to cut metal sheets, using a specific blade. Setup time is the same, but the execution times is set to 10 times slower, and stock loading and loading time is 5 times slower. Finally, the jigsaw's precision error is set to 2 times that of wood. Modulo the cost difference, we apply the same process to evaluate the fabrication cost.
}

\subsection{Pareto front extraction}
In Section 4.3.4 of the main paper,
 we propose two methods to speed up the Pareto front extraction.
 This section provides more details of the two methods.
 For an atomic \enode which has not been previously optimized,
 we simply try a maximum, $P$, different orders of cuts
 (each cutting order is evaluated using the cost metrics in \autoref{subsec:costs}),
 then select the cutting order with minimum precision error ($f_p$)
 and the one with minimum fabrication time ($f_t$).

To apply the branch and bound technique, we need to define the lower and upper bound for the precision error ($f_p$) and the time cost ($f_t$). A term $\mathcal{T}$'s cutting order can be initialized by taking the cutting order used in its contained atomic \enode.  We first select the cutting order with minimized precision of each atomic \enode, then evaluate its precision error defined as the precision upper bound of the term. The time upper bound is set by taking the cutting order with minimized time of each atomic \enode. 
\change{
The precision (time) lower bound is defined as the evaluated precision (time) cost,
ignoring the dependency relationship between cuts belongs to the same stock. }
The dependency relationship indicates the order of a cut with respect to other cuts.
In other words, the time lower bound is computed by the assumption that each cut is independent of the other.

If the precision (or) lower bound is not dominated by the Pareto front of all computed solutions $\sols$, we run an optimization that uses the upper bound as a starting point.  For the optimization, we randomly flip the order of some cuts until $t$ iterations. We set $t$ as 20 in our experiments.



\section{Results and Discussion}

\change{In this section, we provide additional results of our ICEE pipeline for exploring the space of design variations and fabrication plans of Figure 7. \autoref{tab:hv} shows detailed comparison between our pipeline and the no design exploration pipeline. At the end of this supplemental material, for each model, we demonstrate some representative design variations and fabrication plans extracted with our ICEE pipeline and the baseline method.}

\begin{table*}[]
    \centering
\begin{tabular}{l|c|c|c|c|c|c|c|c|c}
\hline
    \textbf{Model} & \textbf{$HV_{base}$} & \textbf{$HV_{our}$} & \textbf{Ref. Point} 
    & \textbf{$M_{base} (\$)$} & \textbf{$P_{base} (\inchsign)$} & \textbf{$T_{base} (m)$}  
    & \textbf{$M_{our} (\$)$} & \textbf{$P_{our} (\inchsign)$} & \textbf{$T_{our} (m)$} \\ \hline
Frame      & 881570   & \textbf{902017}   & (100.0,100.0)       & 10.00 & N/A  & 1.98  & \textbf{8.50}  & N/A  & \textbf{1.37}  \\ \hline
L-Frame    & 796637   & \textbf{797655}   & (100.0,100.0)       & 18.50 & N/A  & 2.15  & 18.50          & N/A  & \textbf{2.05}  \\ \hline
A-bookcase & 698414   & \textbf{702256}   & (100.0,100.0,100.0) & 23.00 & 0.16 & 8.82  & 23.00          & 0.13 & \textbf{8.53}  \\ \hline
S-Chair      & 683887   & \textbf{698016}   & (100.0,100.0)       & 25.50 & N/A  & 7.92  & 25.50          & N/A  & \textbf{5.92}  \\ \hline
Table      & 833919   & \textbf{849705}   & (100.0,100.0)       & 20.00 & N/A  & 7.18  & 20.00          & N/A  & \textbf{5.78}  \\ \hline
F-Cube     & 817648   & \textbf{825935}   & (100.0,100.0)       & 15.50 & N/A  & 3.08  & 15.50          & N/A  & \textbf{2.12}  \\ \hline
Window     & 754863   & \textbf{777711}   & (100.0,100.0)       & 20.00 & N/A  & 4.93  & 20.00          & N/A  & \textbf{2.38}  \\ \hline
Bench      & 355824   & \textbf{369260}   & (100.0,100.0)       & 55.50 & N/A  & 17.03 & \textbf{53.00} & N/A  & 20.38          \\ \hline
A-Chair    & 596346   & 578014            & (100.0,100.0)       & 35.50 & N/A  & 6.25  & 35.50          & N/A  & 9.52           \\ \hline
F-Pot      & 24155184 & \textbf{24264553} & (300.0,300.0,300.0) & 13.00 & 0.29 & 19.13 & 13.00          & 0.26 & \textbf{17.91} \\ \hline
Z-Table    & 19716012 & \textbf{20386788} & (300.0,300.0,300.0) & 55.50 & 0.35 & 30.69 & \textbf{53.00} & 0.28 & \textbf{24.28} \\ \hline
Loom       & 20469248 & \textbf{21040283} & (300.0,300.0,300.0) & 27.50 & 1.47 & 48.26 & \textbf{25.50} & 0.58 & \textbf{43.86} \\ \hline
J-Gym      & 13925991 & \textbf{15657249} & (300.0,300.0,300.0) & 96.00 & 1.82 & 72.02 & 89.00          & 0.62 & \textbf{51.38} \\ \hline
D-Chair    & 543771   & 539241            & (100.0,100.0)       & 35.50 & N/A  & 15.00 & 35.50          & N/A  & 16.20          \\ \hline
Bookcase   & 20266354 & \textbf{21957242} & (300.0,300.0,300.0) & 40.00 & 0.86 & 37.53 & \textbf{30.00} & 0.27 & \textbf{24.71} \\ \hline
Dresser    & 21237296 & \textbf{22866170} & (300.0,300.0,300.0) & 30.00 & 0.61 & 36.48 & 30.00          & 0.14 & \textbf{15.42} \\ \hline

\end{tabular}
 \caption{
 \change{
 Performance comparison between our method and the baseline method (no design exploration pipeline). In this table, we first report the hypervolume value of our method ($HV_{our}$) and the baseline method ($HV_{base}$). The reference points are also listed  used for the hyper-volume computation. We also report the minimal material usage, the minimal cutting precision and the minimal fabrication time of the Pareto fronts of baseline method ($M_{base}, P_{base}, T_{base}$) and the Pareto fronts of our pipeline ($M_{our}, P_{our}, T_{our}$). The cutting precision of some models is not reported for that precision cost has not been taken into the objective metrics.
 We use bold font to indicate the hypervolume and these fabrication costs where we are better than the baseline method.}
 }
    \label{tab:hv}
\end{table*}

\subsection{Carpentry Compiler Parameters}
\change{
During the comparison between our pipeline and the Carpentry Compiler pipeline~\cite{wu2019carpentry}, we use the default parameter setting used in their experiments, the number of \textit{Traversal} $T = 50$, the number of top e-nodes $n = 10$, the maximal different orders of cuts $P = 25$, the population size during E-graph extraction as $120$, the probability of crossover and mutation $p_c=0.95, p_m=0.1$.}

\subsection{Running Time Analysis}
\change{
On average, our pipeline spends 4.2\% of its runtime in design extraction, 5.8\% of its time generating packings, 64.5\% of its time assigning cutting orders, and 21.6\% of its time in applying genetic algorithms to extract Pareto front of terms. This translates to 10.0\% of its runtime in the expansion and contraction phase, 86.1\% in the extraction phase, and the other 3.9\% everywhere else.}

\subsection{Comparison with baseline}


\begin{figure}[h!]
    \centering
    \includegraphics[width = .95\linewidth]{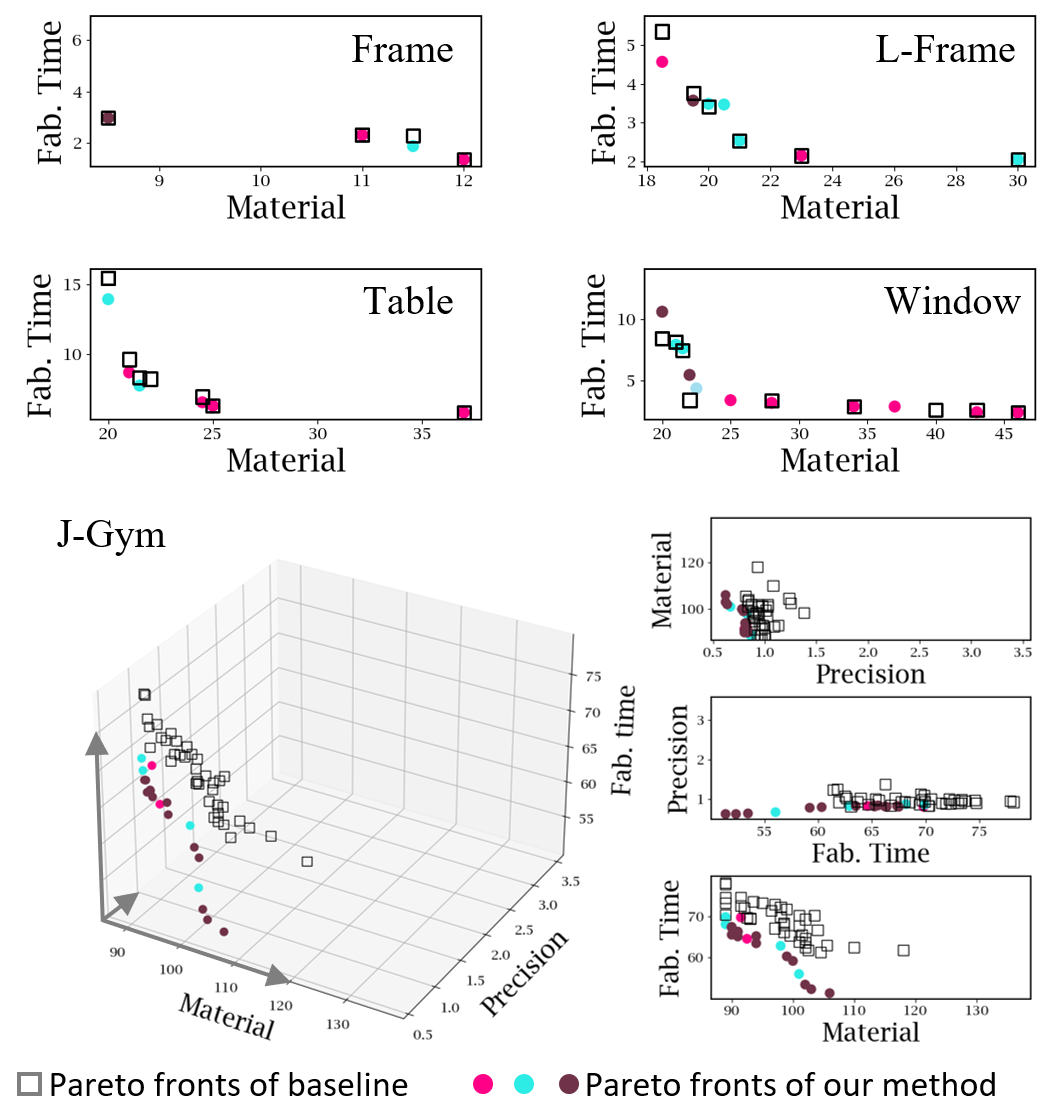}
    \caption{\change{Pareto fronts generated from our pipeline and baseline method.}}
    \label{fig:baseline}
\end{figure}

\begin{table}[]
\begin{tabular}{c|c|c|c}
\hline
 \textbf{Model} & \textbf{$HV_{base}$} & \textbf{$HV_{our}$} & \textbf{Ref. Point}  \\ \hline
Frame   & 901997   & 902017   & (100.0,100.0)       \\ \hline
J-Gym   & 15036794 & 15657249 & (300.0,300.0,300.0) \\ \hline
L-Frame & 797573   & 797655   & (100.0,100.0)       \\ \hline
Table   & 751206   & 849705   & (100.0,100.0)       \\ \hline
Window  & 778158   & 777711   & (100.0,100.0)      \\ \hline
\end{tabular}
    \caption{\change{Hypervolume result of the performance validation experiment. In this table, we report the hypervolume of the Pareto fronts computed from our method ($HV_{our}$) and the "baseline" method ($HV_{base}$) . "Baseline" indicates extracting the Pareto front fabrication plans fro each design variation explored by our method independently with the Carpentry Compiler pipeline~\cite{wu2019carpentry}. The reference points used for hypervolume computation are also reported.}}
    \label{tab:baseline}
\end{table}

\change{
As shown in~\autoref{fig:baseline}, 
we cannot find the exact same front due to convergence
and stochasticity. 
In order to make this comparison, we have tuned
the parameters to be as close as possible.
This means we do not find the exact same hypervolume indicated in~\autoref{tab:baseline}.
However, hypervolume is unintuitive to compare,
so we believe presenting plots for comparison 
is far more compelling.
}


\begin{figure}
    \centering
    \includegraphics[width = \linewidth]{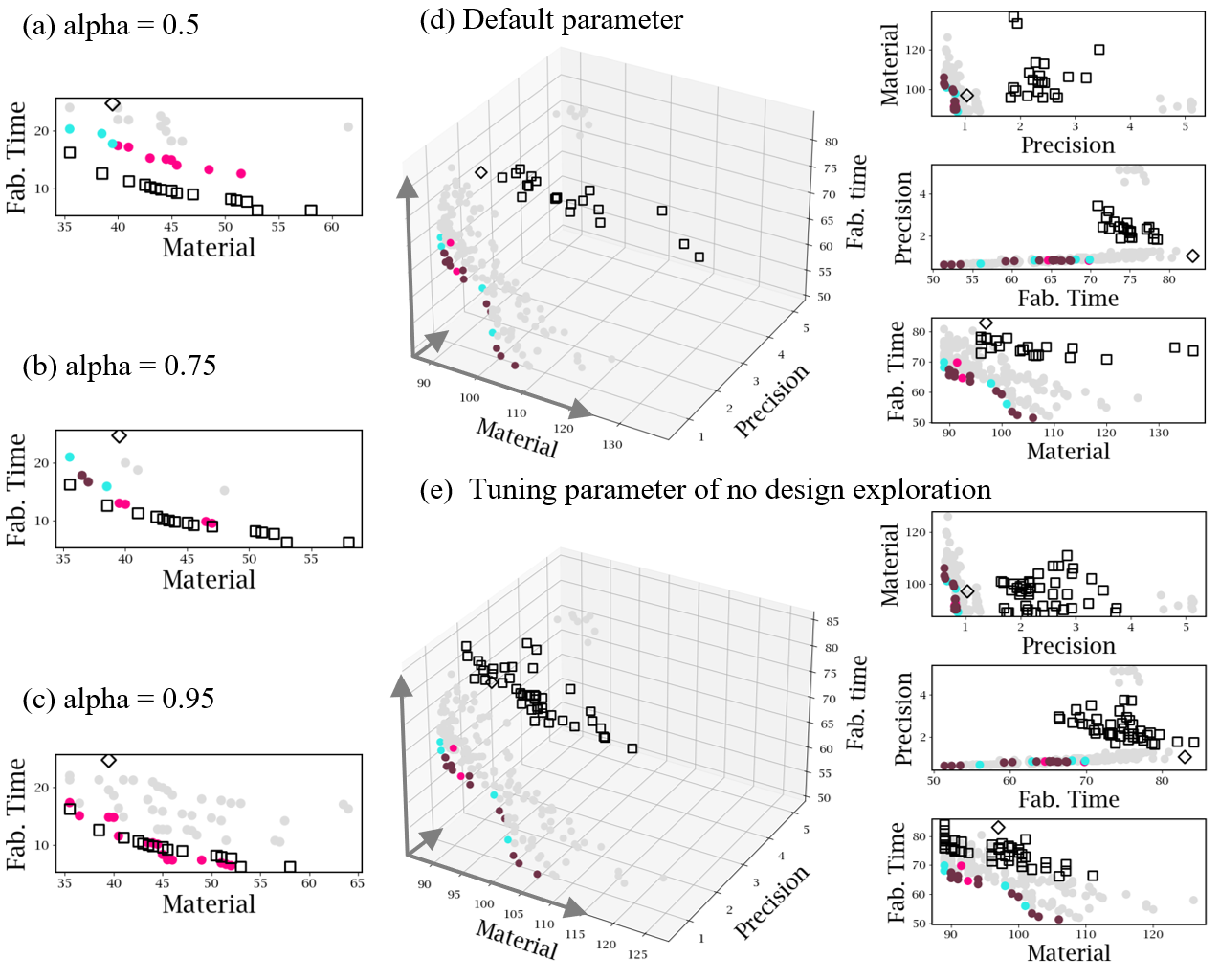}
    \caption{
    \change{
 Pareto fronts of the A-Chair model with different alpha, (a) shows the result of a small alpha ($0.5$), (b) shows the result of the default alpha ($0.75$), (c) shows the result of a bigger alpha ($0.95$). The Pareto fronts generated from no design exploration pipeline stay the same in (a,b,c).  
 Pareto fronts of the J-Gym model with tuning parameters of the no design exploration pipeline, (d) show results of the default parameters, (e) show the Pareto fronts with tuning parameters of the no design exploration pipeline. The Pareto fronts generated from our pipeline stay the same in (d,e). 
    }
    }
    \label{fig:jungle-gym-design-opt-comparison}
\end{figure}

\subsection{Convergence}

\change{
Due to the combinatorial nature of the problem, we cannot
guarantee that we discover the true Pareto front.
In what follows we discuss how this limitation can affect the results of our evaluation.
}



\subsubsection{Increasing the Design Space}
\begin{figure}[h!]
    \centering
    \includegraphics[width = .60\linewidth]{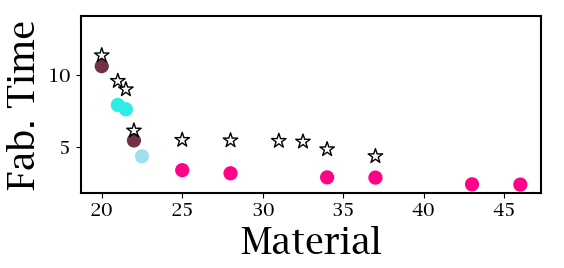}
    \caption{Example showing a limitation of our approach:
    the star dots indicate the Pareto fronts with
    allowing the Window model to have arbitrarily angled connectors; 
    the colored dots indicate ones without arbitrarily angled connectors.
    Both results are generated with the default parameter of our ICEE pipeline.
    }
    \label{fig:lim}
\end{figure}

\change{An implication of the intractable search is that
we have a finite amount of resources to spend searching the space.
Realistically, this means we can only explore a subset of the design space.
Consider the design of the
 ``Window''
 in Figure 7.
If we permit
 this model to have  angled
 joints instead of just 90 degree joints, 
 the space of design variants becomes much larger.
Ideally, the result for this new window variant should be at least as
 optimal as the result for the original window.
However, our approach was able to explore only a
 portion of the vast design space ($\nd=1520052$, compared to $\nd=10463$ for the simpler window model) and found a worse landscape of solutions (\autoref{fig:lim}).
For models with a large design space,
 if the initial design variants picked by our
 algorithm are too far
 from the optimal ones,
 then our algorithm
 may never reach those optimal points.}
 
\change{
To avoid similar potential pitfalls, 
we might be able to allocate our resources 
to effectively direct the search.
In particular, one thing we can do to 
navigate this space more precisely is to expose a parameter, alpha.
}

\change{
$\alpha$ is a parameter that roughly defines the tradeoff between breadth 
(exploring many designs) and depth (exploring variations within a small 
number of designs). For our experiments, we set $\alpha$ to a default
value of $0.75$. However, in cases where the initial design is believed to be 
optimal or close to optimal, the user may increase $\alpha$ to encourage
deeper exploration of the initial and similar designs. In cases 
where our baseline of optimizing the fabrication for a single design
outperforms exploring many designs in the default configuration,
we have found that increasing $\alpha$ to $0.95$ makes the performance 
at least as good again.
}

\change{This note about parameter tuning reveals
interesting implications. 
In particular, our new algorithm does not
always fully dominate fabrication-plan-only exploration
with default parameters.
There are models where the input design is already fairly
optimized and doesn't offer much opportunity for improvement.
Take a model that is simple or is composed of many instances
of the same shape. The Adirondack chair is such an example. 
It would be clear that an optimal packing
arrangement would minimize material, cuts, and error by stacking
cuts; design variations would only deviate from the simplicity of the
fabrication plan and there are no improvements possible.
The baseline method has the advantage with these models,
because it spends its search time deeply exploring
fabrication plans, while ICEE
must also spend time searching across design variations.
Tuning the parameter alpha can help us perform
as well as the baseline method, as seen in~\autoref{fig:jungle-gym-design-opt-comparison} (a) (b) (c).

From these examples, we see that comparing approaches 
while using parameter tuning is trickier and subtler.
}

\subsubsection{Comparisons with~\cite{wu2019carpentry}}

\change{Notably, comparisons with the approaches not 
employing design exploration are also susceptible to the
imprecision in convergence.
In previous sections, we ran each tool with its default
parameters. Tuning the parameters for each approach
would change the results and
make the difference in Pareto fronts difficult to assess.
Neither approach discovers the true Pareto front,
so an absolute comparison depends on the input parameters.
Fortunately, we avoid the brunt of this pitfall:
the selling point of our design space exploration
approach is not about eking out marginal improvements over the 
baseline, but about exploring a completely different space.}


\change{
Though we may not search any one 
design as thoroughly when using default parameters,
our approach is able to do something the baseline cannot:
discover optimizations to the input design
when they exist. Consider elaborate models 
where there are many ways to pack and fabricate any design, 
and due to the complexity and number of parts, the designer
has diminishing intuition about how to achieve 
some desired point in the space of tradeoffs. 

An example that illustrates this complexity is
the Jungle Gym model. The model uses a mix of 
1D and 2D wood and has a number of components.
From Figure~\ref{fig:jungle-gym-design-opt-comparison} (d) (e),
we see that the initial design was suboptimal, and just by
exploring several different design variations, we
find a number of plans that save greatly on our three objectives.
Even the expert plan is completely dominated by the pareto front,
indicating that good designs are not always readily apparent.
}

\change{
Our tool takes much longer to finish running
as it explores the design space jointly with the
fabrication space. Because this is a synthesis problem,
one might ask if allowing the baseline to run for longer
would result in better fabrication plans for the initial 
design. 
We chose not to attempt a comparison of the two 
algorithms for two reasons. Although it is possible
that allowing the fabrication-space-only algorithm
to run longer will help it find lower-cost plans,
there is no parameter that directly controls
how long the algorithm runs. The second reason is that
we are supplying a new approach that explores a different
solution space, and targets a different problem.
Controlling for the running time of the algorithm
would still not create an apples-to-apples comparison.
}

\begin{table}[ht]
    \centering
    \begin{tabular}{c|rrrrrrrrr}
    \hline
        \multicolumn{9}{c}{Percentage improvement (\%)}\\
        \hline
        \multirow{2}{*}{Model} & \multicolumn{8}{c}{Carpenter price ($\$/hour$)}\\
        & 0 & 10 & 20 & 40 & 80 & 160 & 240 & 400\\
        \hline
Frame      & 15 & 19 & 21 & 20 & 15 & 10 & 12 & 16  \\
L-Frame    & 0  & 1  & 2  & 2  & 1  & 3  & 1  & 0   \\
A-bookcase & 0  & 1  & 2  & 6  & 7  & 6  & 6  & 5   \\
S-Chair    & 0  & 1  & 2  & 2  & 6  & 9  & 11 & 14  \\
Table      & 0  & 1  & 2  & 3  & 4  & 7  & 10 & 12  \\
F-Cube     & 0  & 4  & 5  & 3  & 3  & 5  & 6  & 8   \\
Window   & 0  & 2  & 4  & 12 & 20 & 26 & 28 & 31  \\
Bench      & 5  & 4  & 4  & 4  & 4  & 5  & 1  & -4  \\
A-Chair    & 0  & -2 & -4 & -4 & -3 & -4 & -9 & -16 \\
F-Pot      & 0  & 3  & 4  & 5  & 7  & 9  & 9  & 8   \\
Z-Table    & 5  & 5  & 6  & 8  & 11 & 11 & 11 & 12  \\
Loom       & 7  & 3  & 1  & 0  & 3  & 3  & 4  & 5   \\
J-Gym      & 7  & 7  & 7  & 8  & 11 & 17 & 20 & 23  \\
D-Chair    & 0  & 1  & 1  & 2  & 3  & 2  & 0  & -2  \\
Bookcase   & 25 & 16 & 10 & 7  & 8  & 12 & 14 & 18  \\
Dresser    & 0  & 2  & 4  & 6  & 8  & 25 & 33 & 42 
    \end{tabular}
    \caption{\change{The percent improvement of minimum-cost (after scalarization) plans of the baseline compared to the minimum-cost plans of design space exploration, when scalarized at different prices.}}
    \label{tab:scalarization}
\end{table}

\subsection{Scalarization Results}

\change{
\autoref{tab:scalarization} shows the scalarization of some of the tradeoffs
found on the Pareto front with our method.
Note that these results are all for the wood models with our default cost metric.
When time isn't worth a lot, plans with low material cost dominate.
When time is worth more much than materials, low time become cheapest.
We observe a variety of fabrication plans being effective at different points.}

\begin{figure*}
\centering
\includegraphics[width=0.85\linewidth]{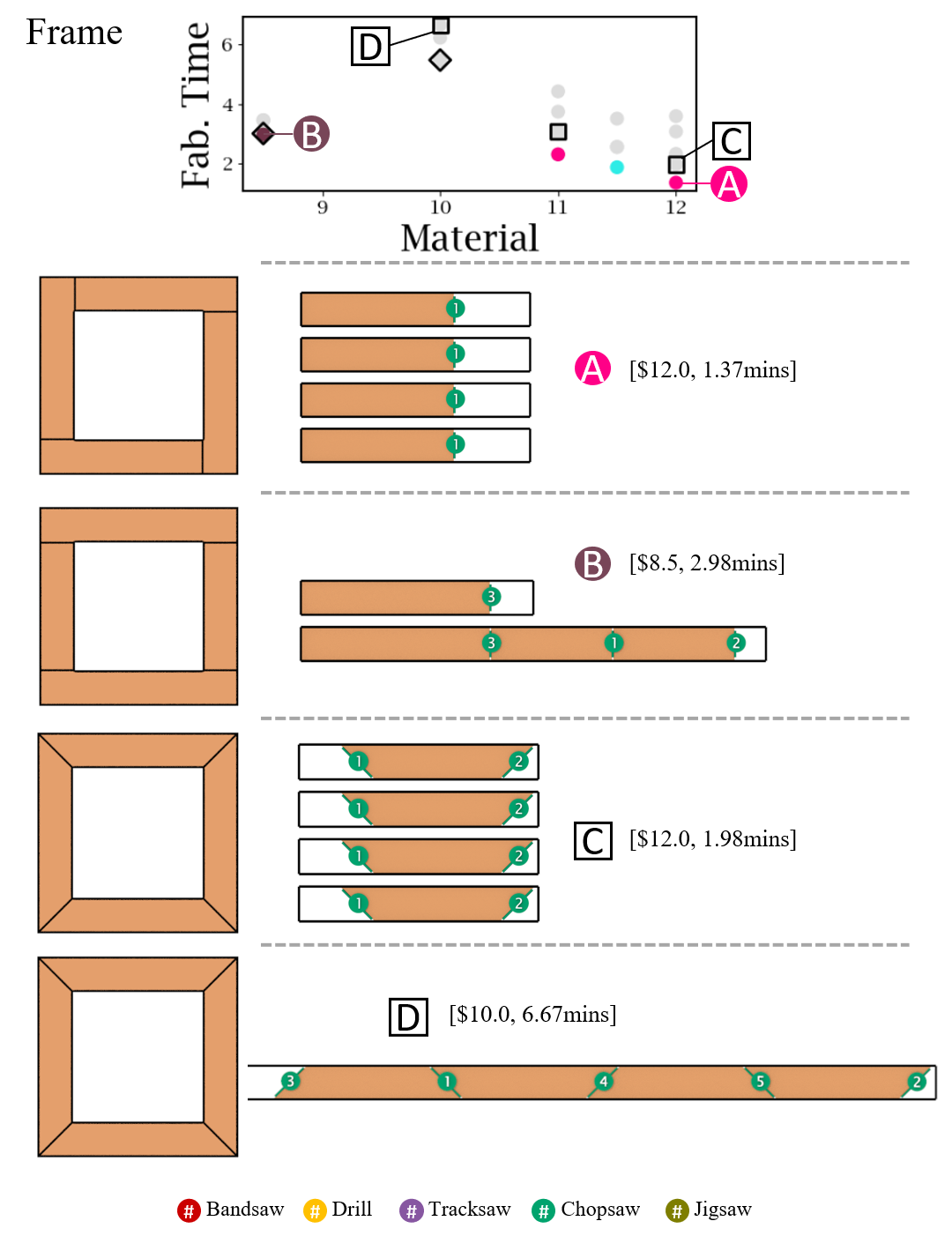}
\caption{Design variants and fabrication plans extracted with our ICEE algorithm (Frame)}
\label{fabplan}
\end{figure*}

\begin{figure*}
\centering
\includegraphics[width=0.85\linewidth]{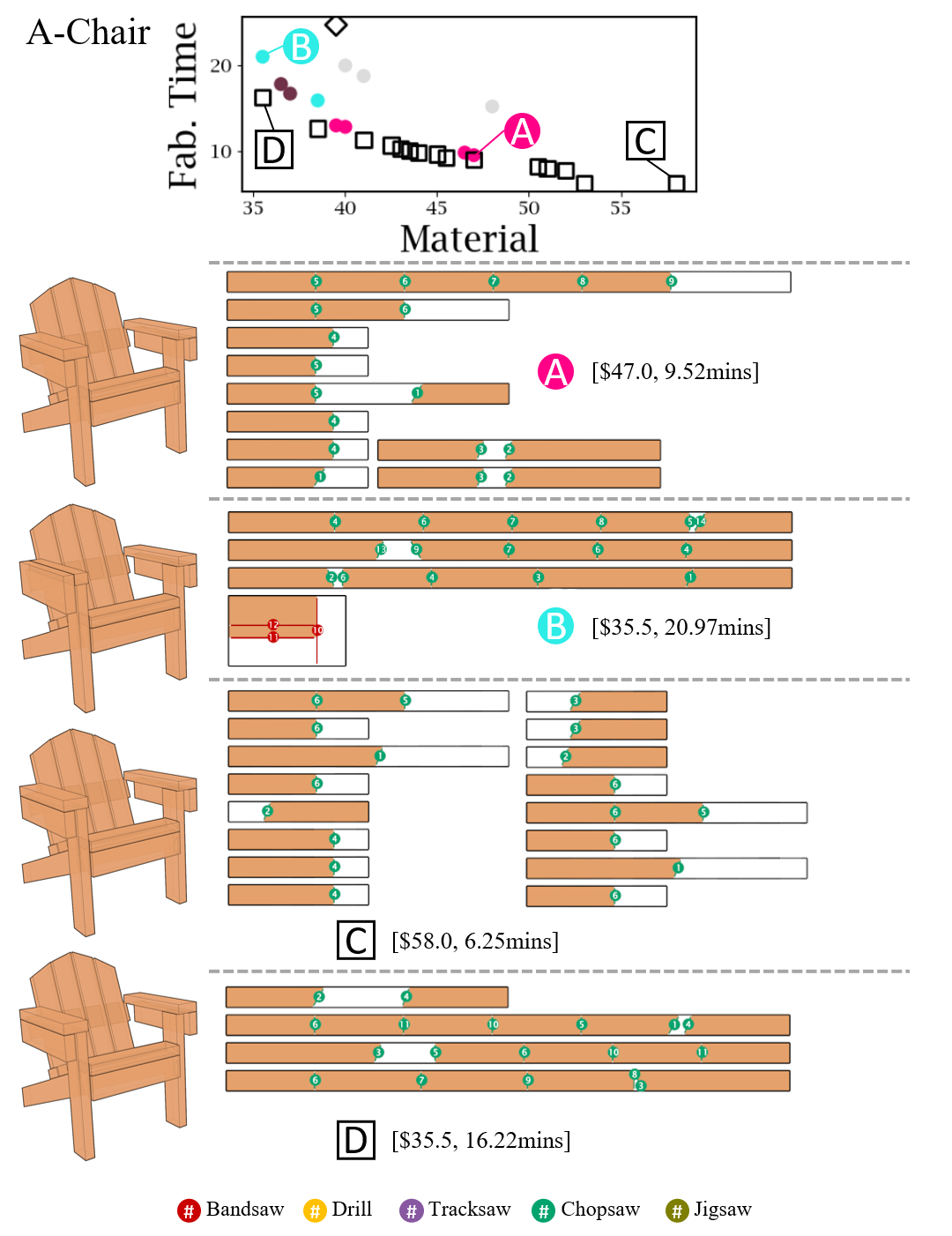}
\caption{Design variants and fabrication plans extracted with our ICEE algorithm (A-Chair)}
\label{fabplan}
\end{figure*}

\begin{figure*}
\centering
\includegraphics[width=0.85\linewidth]{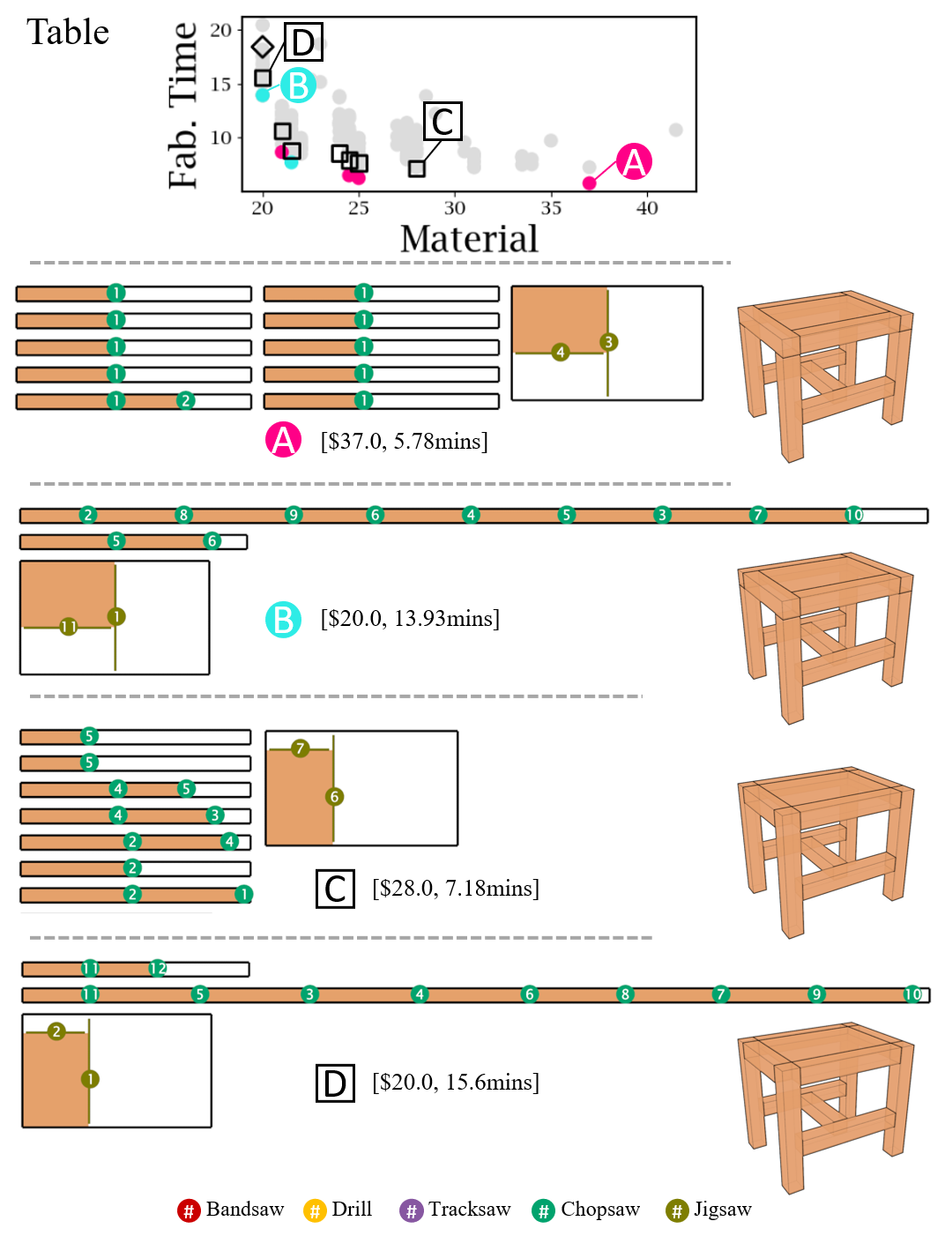}
\caption{Design variants and fabrication plans extracted with our ICEE algorithm (Table)}
\label{fabplan}
\end{figure*}

\begin{figure*}
\centering
\includegraphics[width=0.85\linewidth]{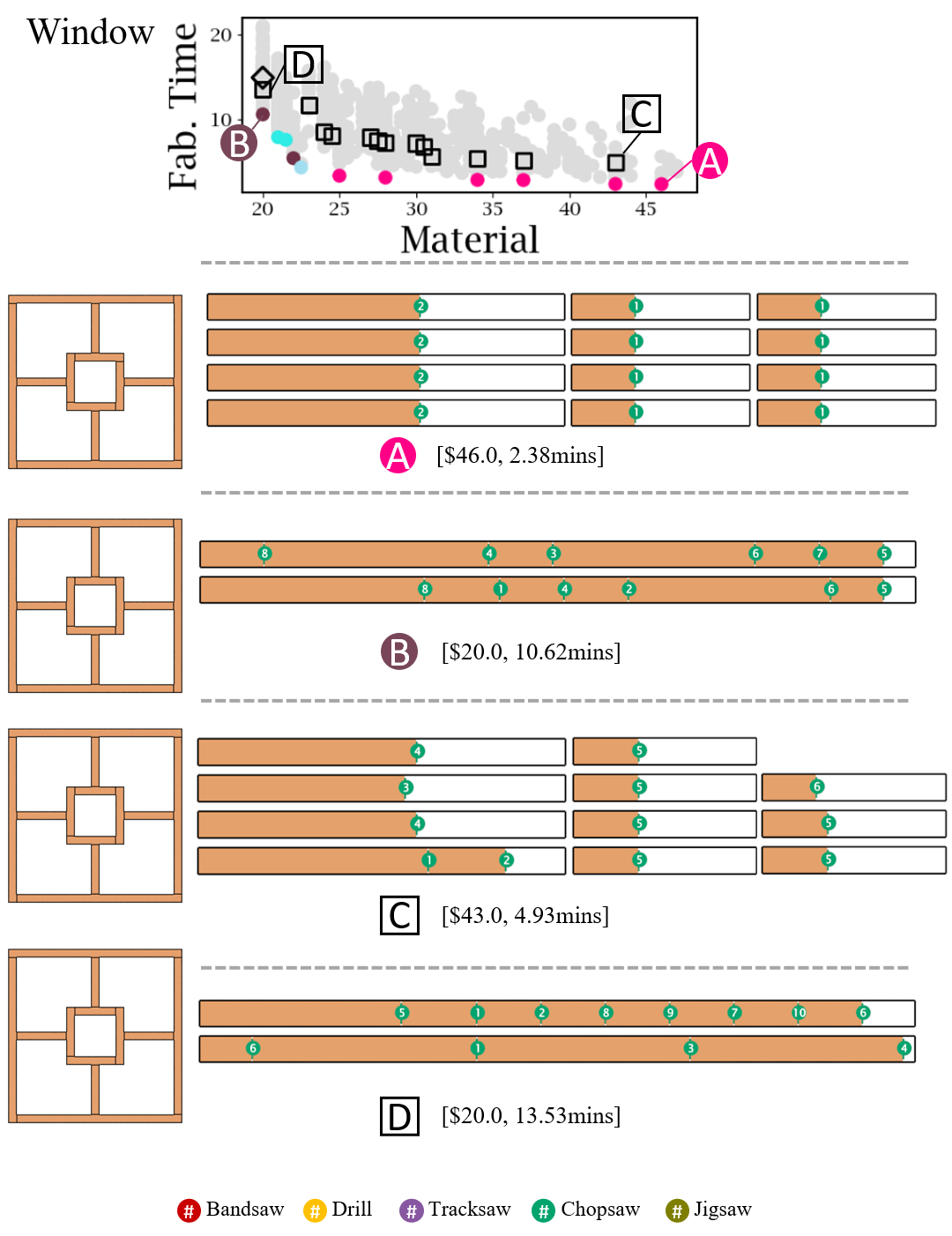}
\caption{Design variants and fabrication plans extracted with our ICEE algorithm (Window)}
\label{fabplan}
\end{figure*}

\begin{figure*}
\centering
\includegraphics[width=0.85\linewidth]{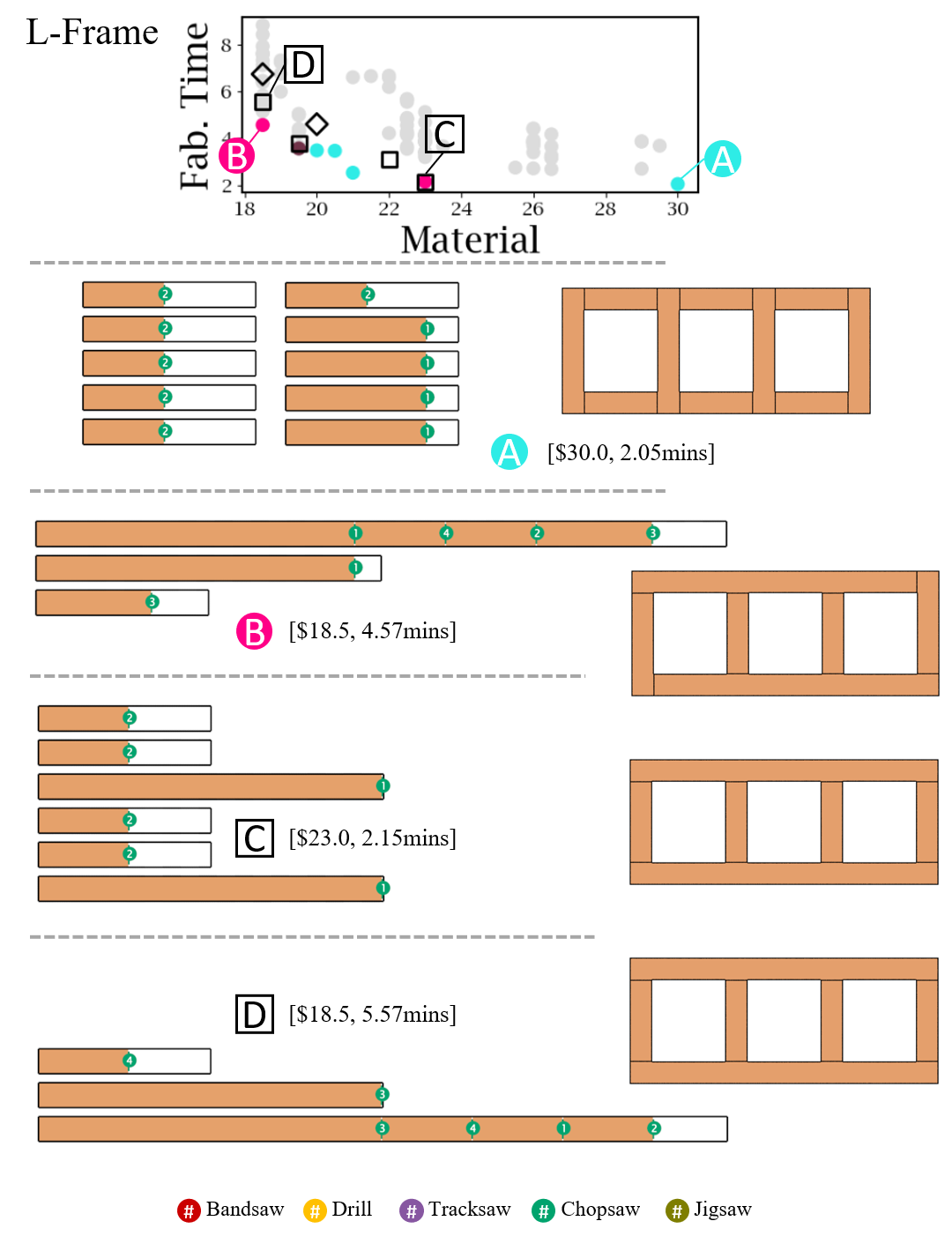}
\caption{Design variants and fabrication plans extracted with our ICEE algorithm (L-Frame)}
\label{fabplan}
\end{figure*}

\begin{figure*}
\centering
\includegraphics[width=0.85\linewidth]{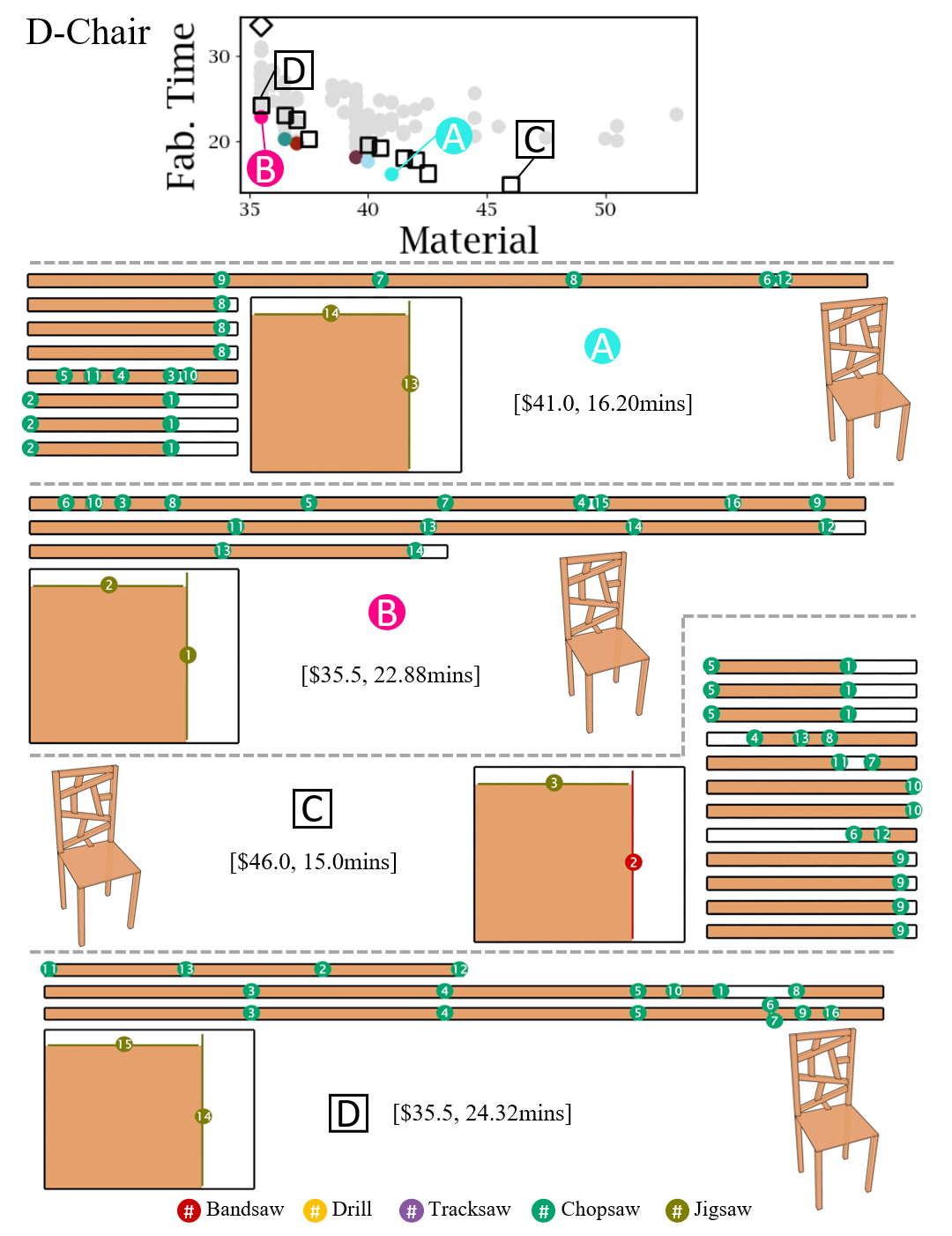}
\caption{Design variants and fabrication plans extracted with our ICEE algorithm (D-Chair)}
\label{fabplan}
\end{figure*}

\begin{figure*}
\centering
\includegraphics[width=0.85\linewidth]{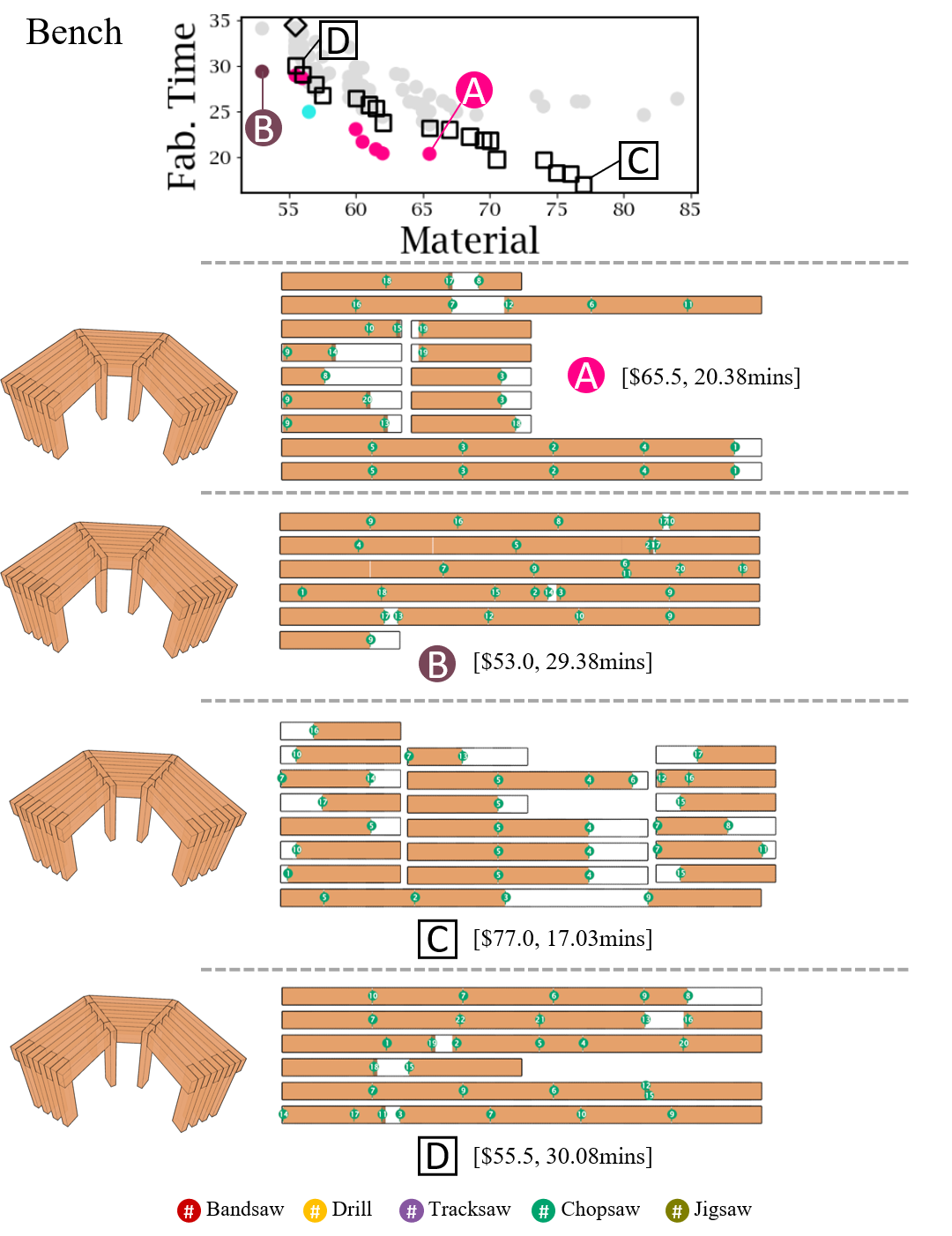}
\caption{Design variants and fabrication plans extracted with our ICEE algorithm (Bench)}
\label{fabplan}
\end{figure*}

\begin{figure*}
\centering
\includegraphics[width=0.85\linewidth]{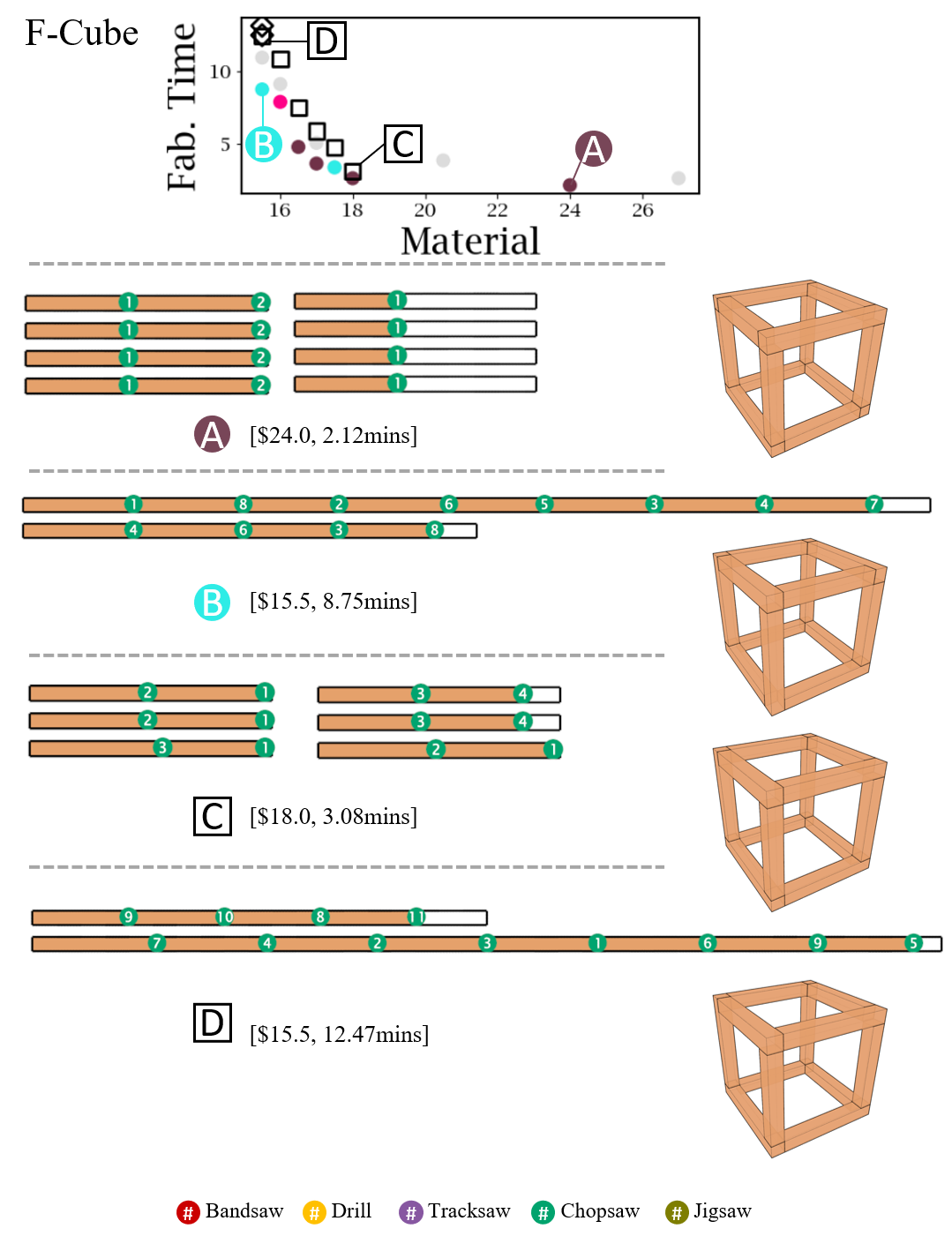}
\caption{Design variants and fabrication plans extracted with our ICEE algorithm (F-Cube)}
\label{fabplan}
\end{figure*}

\begin{figure*}
\centering
\includegraphics[width=0.85\linewidth]{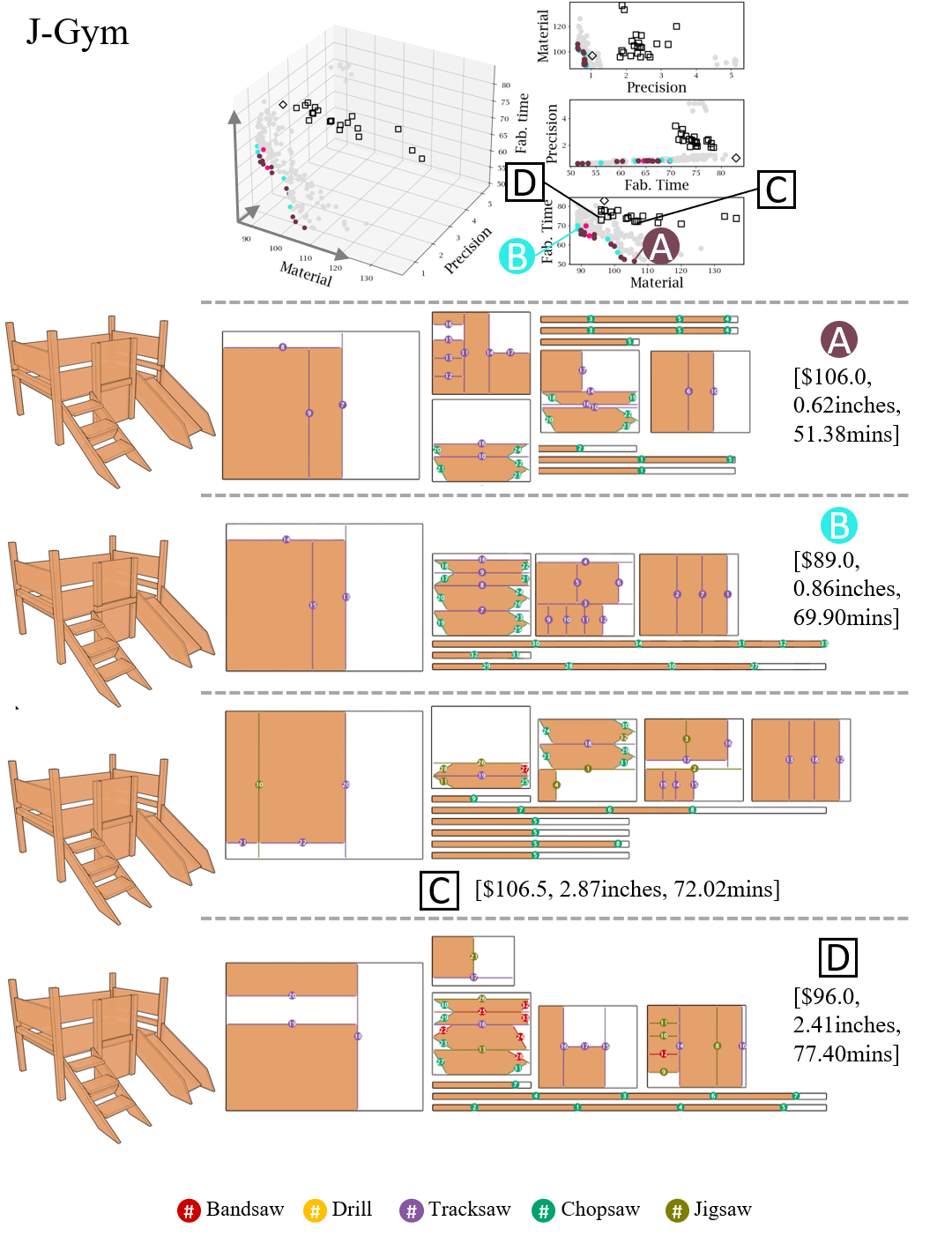}
\caption{Design variants and fabrication plans extracted with our ICEE algorithm (J-Gym)}
\label{fabplan}
\end{figure*}

\begin{figure*}
\centering
\includegraphics[width=0.85\linewidth]{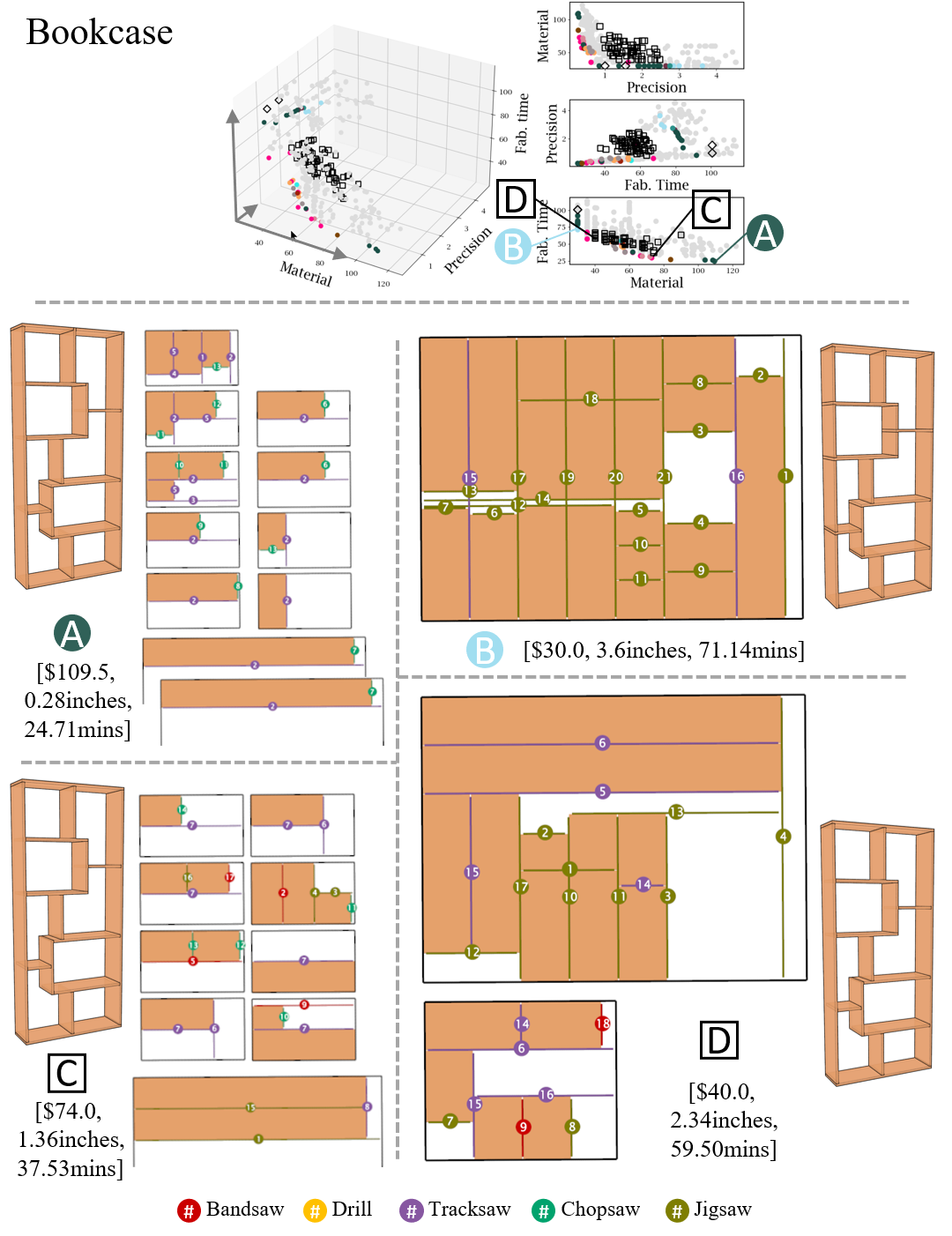}
\caption{Design variants and fabrication plans extracted with our ICEE algorithm (Bookcase)}
\label{fabplan}
\end{figure*}

\begin{figure*}
\centering
\includegraphics[width=0.85\linewidth]{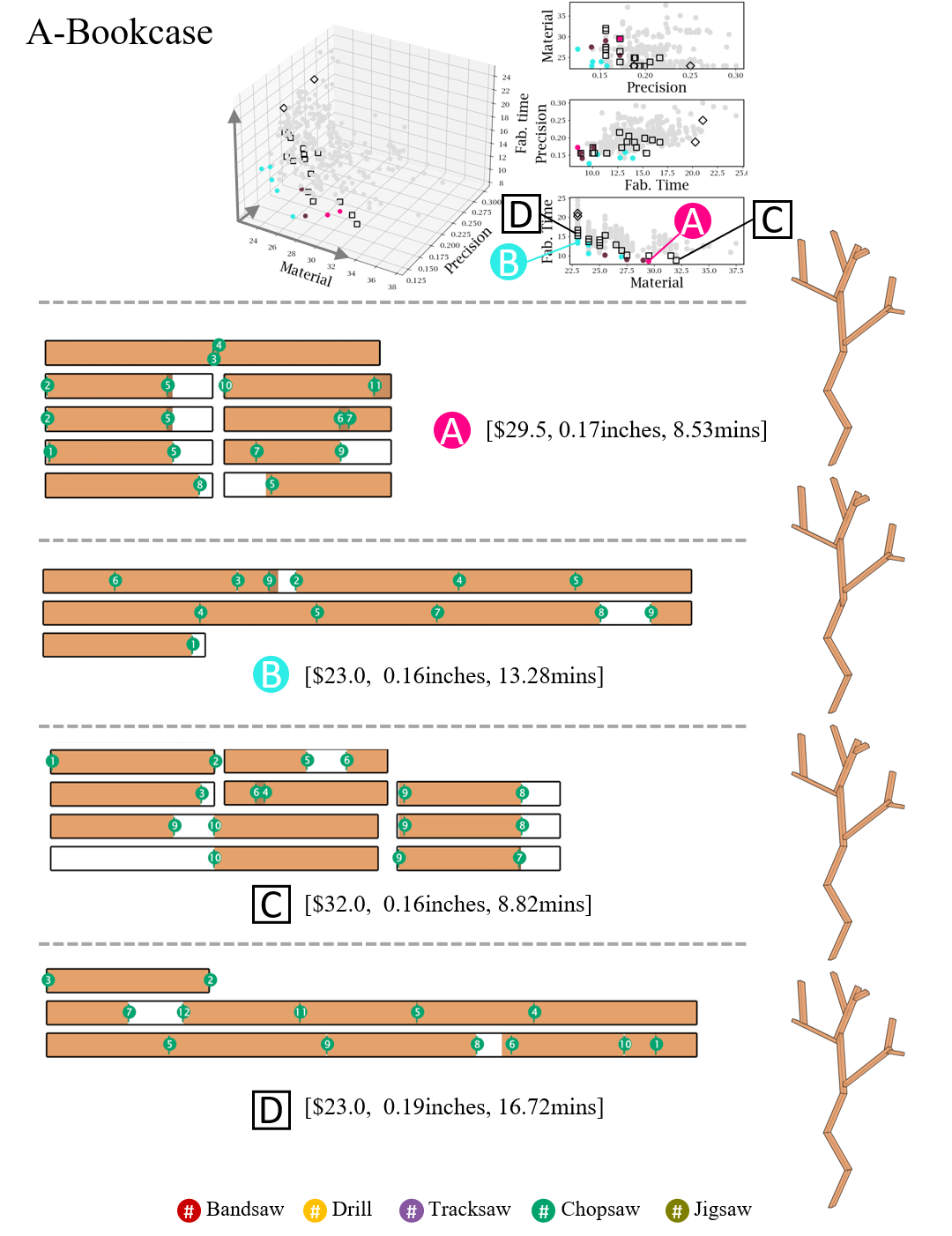}
\caption{Design variants and fabrication plans extracted with our ICEE algorithm (A-Bookcase)}
\label{fabplan}
\end{figure*}

\begin{figure*}
\centering
\includegraphics[width=0.85\linewidth]{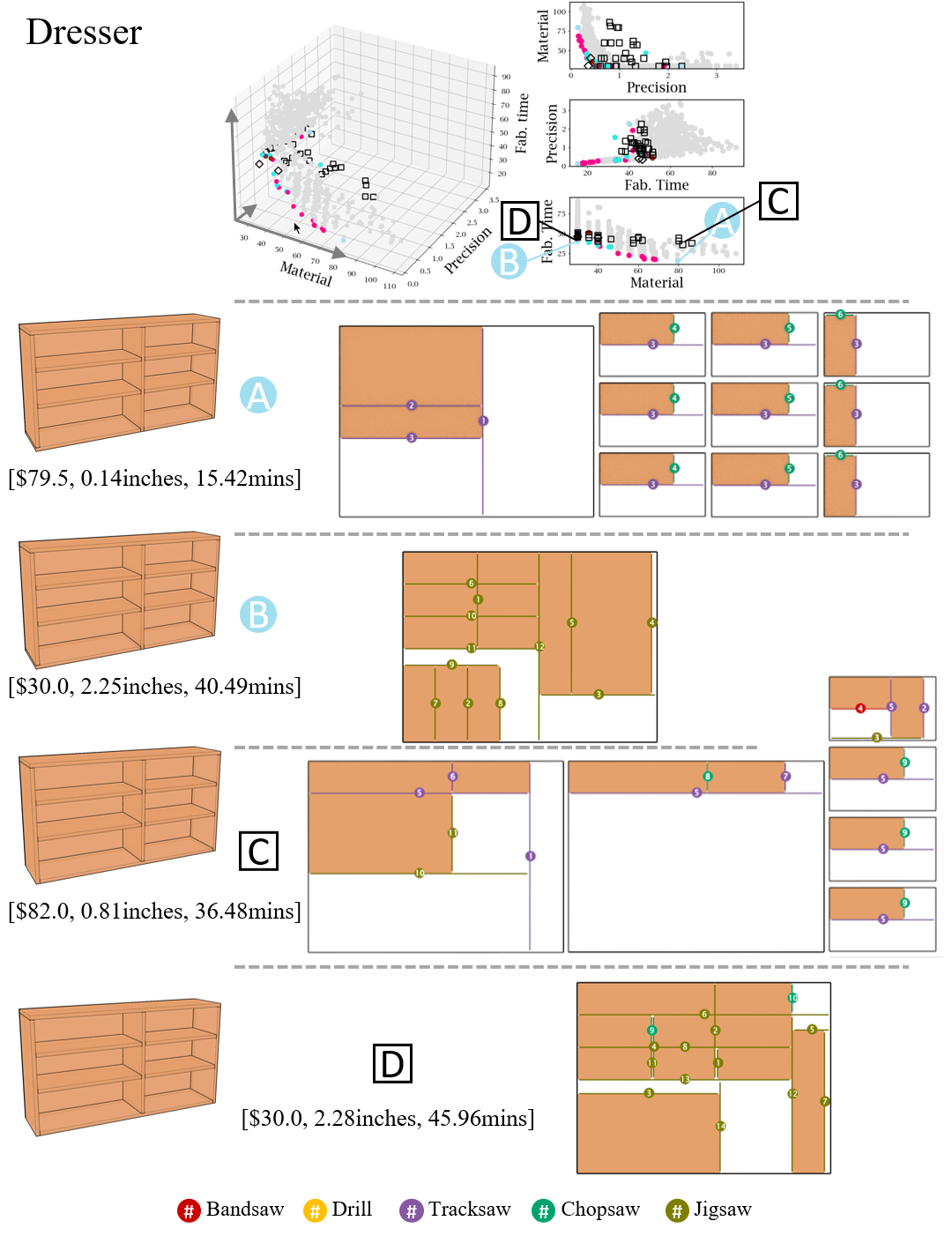}
\caption{Design variants and fabrication plans extracted with our ICEE algorithm (Dresser)}
\label{fabplan}
\end{figure*}

\begin{figure*}
\centering
\includegraphics[width=0.85\linewidth]{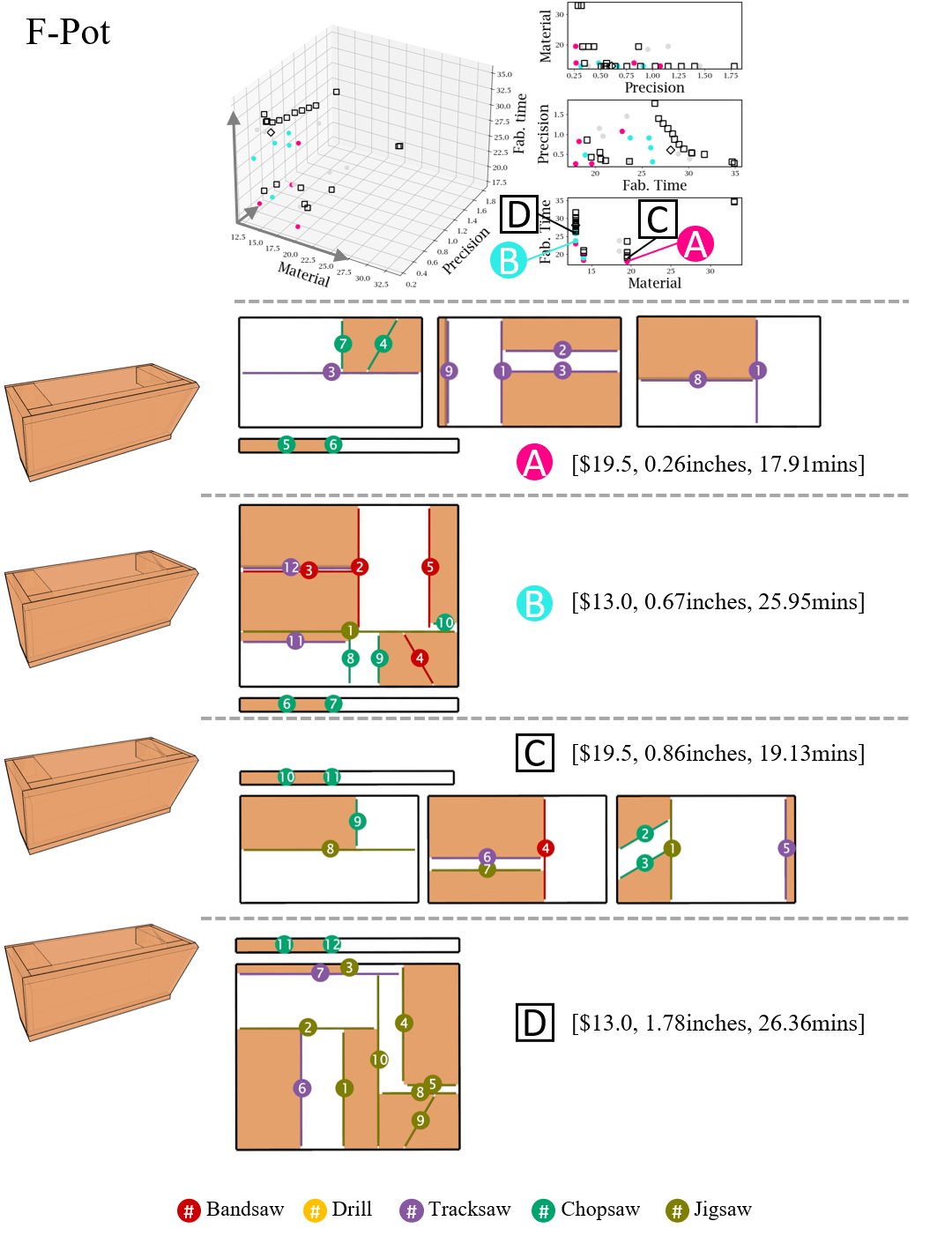}
\caption{Design variants and fabrication plans extracted with our ICEE algorithm (F-Pot)}
\label{fabplan}
\end{figure*}

\begin{figure*}
\centering
\includegraphics[width=0.85\linewidth]{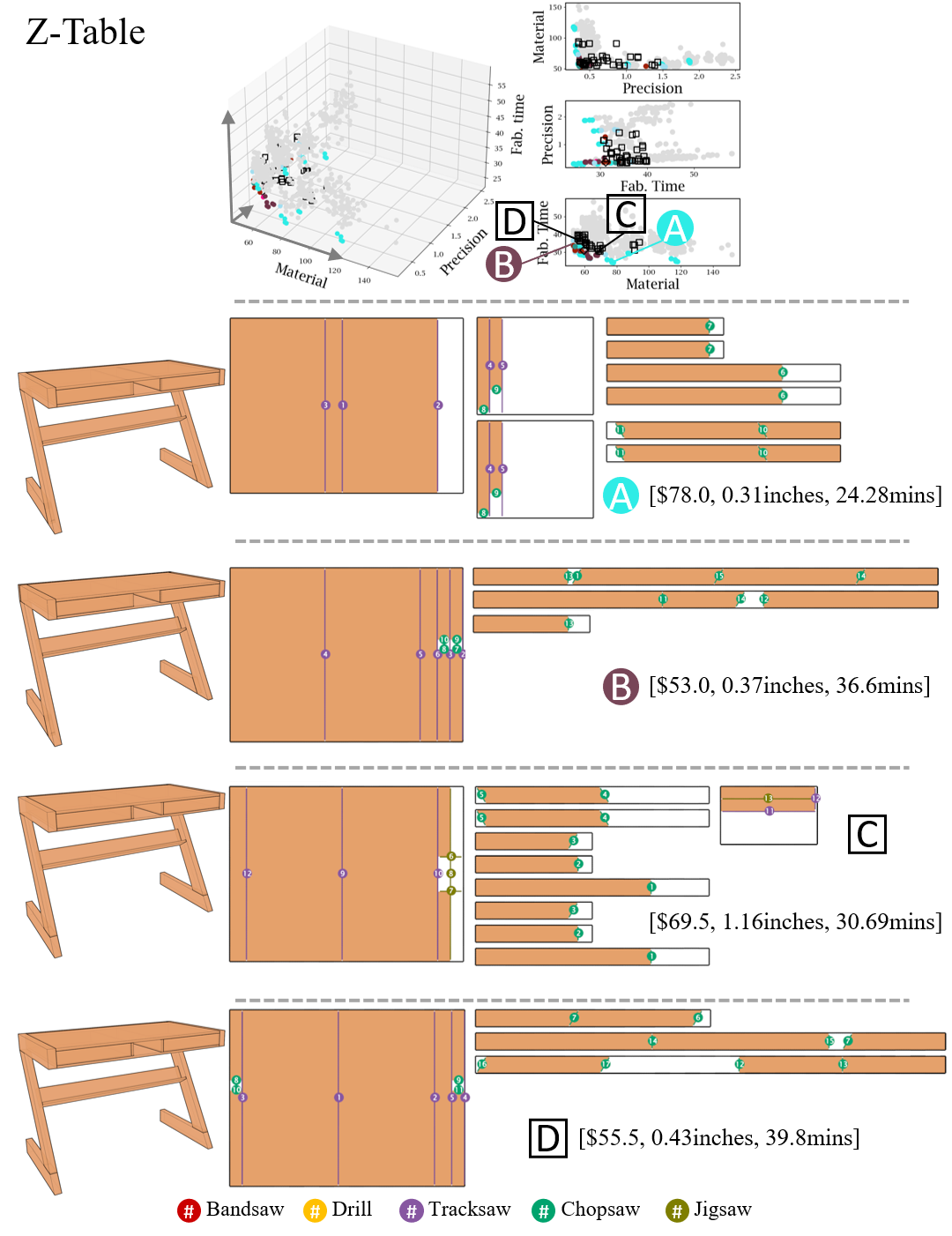}
\caption{Design variants and fabrication plans extracted with our ICEE algorithm (Z-Table)}
\label{fabplan}
\end{figure*}

\begin{figure*}
\centering
\includegraphics[width=0.85\linewidth]{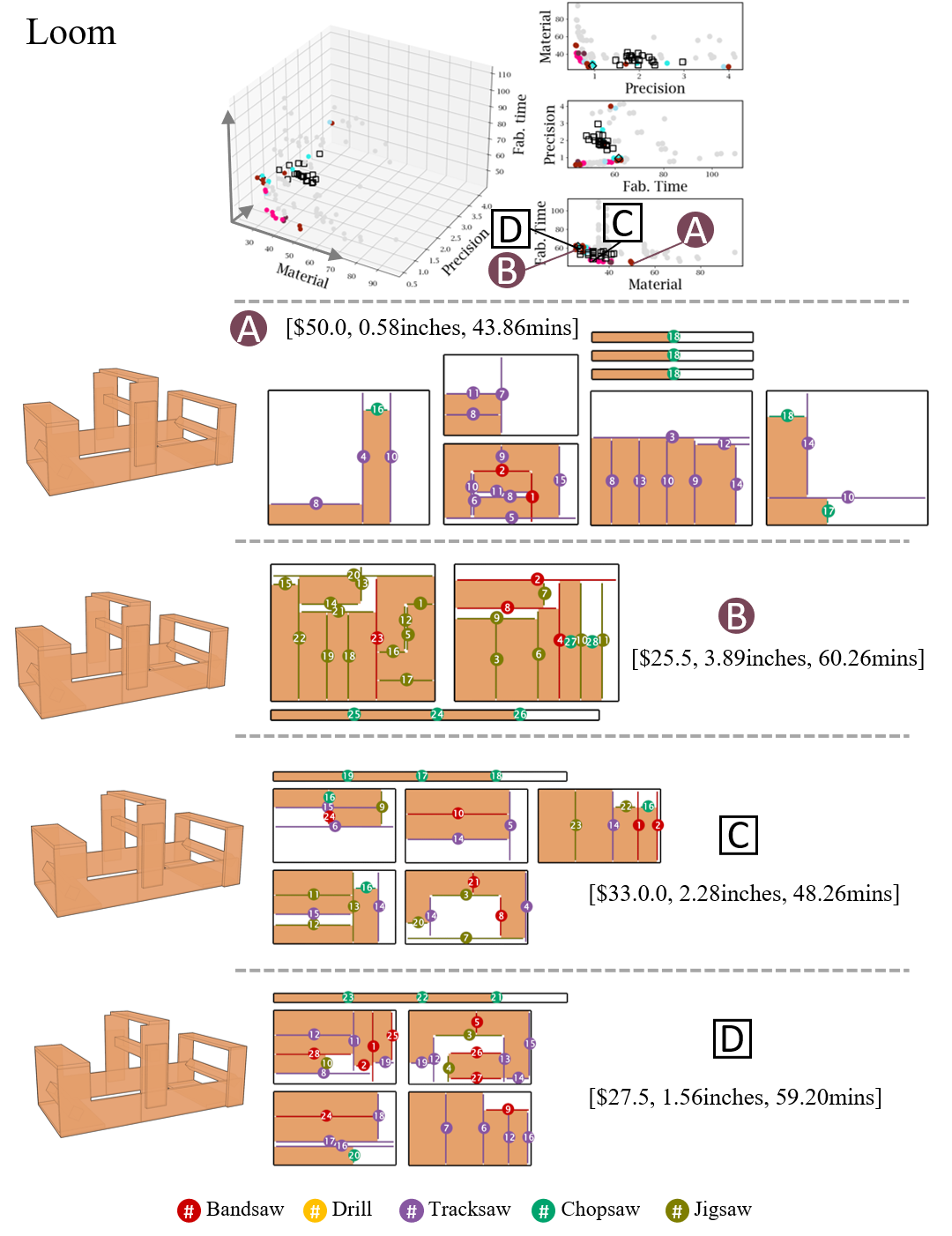}
\caption{Design variants and fabrication plans extracted with our ICEE algorithm (Loom)}
\label{fabplan}
\end{figure*}

\bibliographystyle{ACM-Reference-Format}
\bibliography{sample-bibliography}

\end{document}